\documentclass[journal]{vgtc}                     % final (journal style)
\usepackage{enumitem}
\usepackage{tikz} 
\usetikzlibrary{patterns}
\onlineid{0}
\vgtccategory{Research}
\vgtcpapertype{application/design study}
\title{Guided Exploration of Iterative Schedule Modifications:\\A Design Study on Railway Traction Unit Scheduling}
\author{%
  \authororcid{Andreas Zajic}{0009-0008-4582-8537},
  \authororcid{Vera Hechtl}{0009-0009-7335-2692},
  \authororcid{Cornelia Geischläger}{0009-0006-1174-8462},
  \authororcid{Maximilian Kunovjanek-Bachler}{0000-0002-9111-0525},
  {Thomas Hulka},\\
  {Anna-Lena Penk},
  \authororcid{Belma Turan}{0000-0001-6535-9856},
  \authororcid{Nadine Schwab}{0000-0002-3802-8979},
  \authororcid{Maximilian Viehauser}{0009-0008-7618-1123},
  \authororcid{Helwig Hauser}{0000-0003-0395-3192},
  \authororcid{Kre\v{s}imir Matkovi\'{c}}{0000-0001-9406-8943}
}
\authorfooter{
  \item
  	A.\ Zajic, V.\ Hechtl, T.\ Hulka, A.-L.\ Penk, and Kr.\ Matkovi\'{c} are with VRVis Research.
  	E-mail: \{Zajic | Hechtl | Hulka | Matkovic\}@VRVis.at, anna-lena.penk@web.de.
  \item
  	M.\ Kunovjanek-Bachler \& B.\ Turan are with the Austrian Federal Railways.
  	\item
  	H.\ Hauser is with the Univ.\ of Bergen.
  	E-mail: Helwig.Hauser@UiB.no.
    \item
  	C.\ Geischläger is with dwh GmbH.
  	E-mail: cornelia.geischlaeger@dwh.at.
    \item
  	N.\ Schwab is with the Inst.\ for Information Systems Engineering, TU Wien. 
  	E-mail: nadine.weibrecht@tuwien.ac.at.
    \item
  	M.\ Viehauser is with the Inst.\ of Statistics and Mathematical Methods in Economics, 
    TU Wien. 
  	E-mail: maximilian.viehauser@tuwien.ac.at.
}
\abstract{%
Traction unit scheduling in large railway networks involves complex operational constraints: multi-objective optimization produces feasible circulation plans under ideal assumptions, while simulation is required to assess their robustness under realistic operating conditions.
A critical refinement mechanism relies on crossing operations, in which co-located traction units exchange their remaining schedules to reduce delay propagation.
The space of possible crossing sequences, however, grows exponentially. 
Existing tools provide limited support for identifying promising candidates, evaluating their impact, and managing the resulting exploration. % process.
We present an interactive visual exploration approach that tightly couples schedule visualization, simulation-based evaluation, and a three-level guidance mechanism to support the systematic exploration and interactive optimization of traction unit circulation plans.
The system renders the circulation plan in its domain-familiar form and integrates simulation results to expose delay propagation directly within the planning context.
A three-level guidance framework aggregates crossing candidates spatially and ranks them by estimated impact on key performance indicators (KPIs) at an overview level, while exposing detailed per-candidate evaluation at a detail level to support informed decision-making.
Applying a crossing change triggers an automatic schedule recomputation and re-simulation, with a provenance-based history mechanism enabling the non-linear exploration of alternative modification paths.
We demonstrate the approach through real-world use case scenarios and report substantial reductions in the time and effort required to identify and evaluate promising schedule modifications compared to the current workflow.
}
\graphicspath{{figs/}{figures/}{pictures/}{images/}{./}} % where to search for the images
\usepackage{tabu}                      % only used for the table example
\usepackage{booktabs}                  % only used for the table example
\usepackage{lipsum}                    % used to generate placeholder text
\usepackage{mwe}                       % used to generate placeholder figures
\usepackage{mathptmx}                  % use matching math font
\begin{document}
%
%%%%%%%%%%%%%%%%%%%%%%%%%%%%%%%%%%%%%%%%%%%%%%%%%%%%%%%%%%%%%%%%
%%%%%%%%%%%%%%%%%%%%%% START OF THE PAPER %%%%%%%%%%%%%%%%%%%%%%
\firstsection{Introduction}
\maketitle
The efficient scheduling of traction units to railway services is a fundamental problem in railway operations. 
Traction units are powered railway vehicles that drive scheduled services; they include conventional locomotives hauling separate carriages as well as multiple units, in which propulsion and passenger or cargo accommodation are integrated into a single vehicle.
In large railway networks, traction unit schedules must satisfy numerous operational constraints, including the number and types of available units, the availability of qualified train drivers, maintenance intervals, and dependencies between services. The scale of the problem and the large number of interacting constraints make computing globally optimal traction unit schedules impossible for large-scale railway networks.

In collaboration with a major Austrian railway operator, we study a real-world planning workflow used to manage this complexity. Railway operators typically employ hierarchical planning procedures that refine solutions at several levels of detail. At an early stage, traction units are clustered by region and type, and a coarse seven-day circulation plan is generated by a multi-objective optimization process. The plan is created under the constraint that, for each originating station, the number and types of traction units present at the beginning and end of the planning period must match. This allows weekly plans to be seamlessly concatenated, enabling the construction of long-term schedules by repeating the seven-day plan throughout the year. In this paper, the domain experts and planning professionals who contributed to this topic are listed as co-authors, with professional backgrounds in railway traction unit scheduling. There were no external participants.

While deterministic optimization tools can be used to produce feasible schedules for small networks, they assume ideal operating conditions and do not account for delays or operational disturbances. Stochastic simulation can therefore be applied afterwards to evaluate how the planned schedule performs under realistic operating conditions. By simulating plan execution, potential issues such as delay propagation can be revealed, helping experts to identify parts of the schedule where modifications may improve operational robustness.

An important modification mechanism is the \emph{crossing}, i.e., points in space and time where two operationally equivalent traction units (same type or capability) are simultaneously present at a station and can exchange their remaining schedules. Such crossings can reduce delay propagation, for example when a delayed unit overlaps with another unit that has sufficient slack before its next departure. Although potential crossings can be identified automatically, deciding which crossings to apply and evaluating their effects remains a tedious manual process. Experts must iteratively test alternative crossings and inspect their consequences, making systematic exploration of improvement opportunities inefficient.

The application of a crossing modifies the schedule and may introduce new crossing opportunities. Exploring all possible sequences of crossings would quickly lead to a combinatorial explosion, as the number of possible modification sequences grows exponentially. Therefore, in practice, experts explore modifications incrementally, evaluating crossings that are assumed promising one by one. However, current tools provide limited support to efficiently identify promising modifications and manage the resulting exploration process.

To address this challenge, we present an interactive visual  approach that supports the exploration and refinement of traction unit schedules. We realized this approach in a coordinated multiple views interface that integrates simulation results, enabling experts to analyze the structure and operational behavior of a solution. Users can apply crossings directly in the interface, after which the schedule is recomputed and a new simulation is automatically initiated to evaluate its operational impact. By tightly coupling visualization, modification, and simulation, the approach establishes a central hub in the planning workflow, enabling iterative exploration and comparison of alternative schedule states.

A key element of our approach is a \emph{three-level guidance mechanism} that supports the identification and evaluation of promising schedule modifications. Modification opportunities are aggregated at an overview level to highlight locations where crossings are possible and to show their potential impact. At a detailed level, individual crossing candidates are exposed and ranked according to a selected evaluation metric, while additional contextual attributes are revealed on demand to support expert decision-making. Although developed in the context of railway scheduling, this guidance approach is also applicable to other iterative analysis workflows in which candidate actions can be ranked by certain metrics, but require additional contextual information for informed selection.

Because each modification creates a new schedule state, the system also supports iterative exploration by maintaining a history of explored states. Users can navigate the exploration timeline, return to earlier states when a sequence of modifications proves unproductive, and store promising schedule variants for later comparison. This enables a structured exploration of alternative refinement paths while avoiding an exhaustive combinatorial search.

The main contributions of this paper can be summarized as follows:%
\vspace{-0.5ex}\begin{itemize}\setlength\itemsep{-0.5ex}
\item \textbf{A task characterization for the iterative exploration and refinement of traction unit schedules}, derived in close collaboration of visualization and railway planning experts.
\item \textbf{A three-level visual guidance framework} for iterative modification
workflows in which candidate actions must be grouped, ranked by a primary metric,
and evaluated using secondary contextual attributes.
This progressive disclosure pattern is applicable beyond railway scheduling to other 
domains where fully automatic optimization is infeasible and human expertise drives
solution refinement, as long as a grouping criterion, a primary ranking metric, and
a simulation or evaluation model are available.

\item \textbf{A coordinated multiple views system} that acts as a central hub for schedule exploration, interactive modification, and simulation-based evaluation, enabling iterative refinement through automatic re-simulation and provenance-based exploration of alternative schedule states.
\end{itemize}
Beyond the specific railway application, the core design contributions, the
three-level progressive disclosure guidance pattern and the tight coupling of
interactive modification with automatic simulation and provenance tracking,
address recurring challenges in human-in-the-loop visual analytics for discrete
combinatorial optimization problems, and are intended to inform future work in
this space.
\section{Related Work}
Our work relates to previous research in three main areas: visualization of transportation systems, visual analytics for optimization and planning, and guidance in visualization. 
\subsection{Visualization of Transportation Systems}
Visualization and visual analytics have been applied to the analysis of transportation systems and mobility data.
Andrienko and Andrienko provide a comprehensive overview of visual analytics methods for movement data, describing approaches to exploring spatiotemporal mobility patterns and transportation flows~\cite{Andrienko2013MovementVA}.
Other work focused on revealing structures in large mobility datasets, such as origin--destination flows or trajectory collections~\cite{Guo2009OD,Tominski2012Trajectory}.
These approaches demonstrate how visualization can reveal patterns in complex transportation data and support the analysis of mobility networks.

Several visualization systems specifically address railway operations.
Wunderlich et al.\ presented a design study on visualizing delay uncertainty in train trip planning, combining temporal representations with uncertainty visualization to help users understand how delays affect connections and travel reliability~\cite{Wunderlich2017Delay}.
More recently, Rajendran presented a visual analytics approach for exploring disruptions in railroad networks through coordinated abstract and geographic views~\cite{Rajendran2025RailDisruption}.
These systems visualize static or historical operational data, where the
visualization role is purely analytical: the data is fixed and the user observes.
Our work introduces a fundamentally different coupling: the visualization is
connected to a live simulation backend, so each user-initiated modification
triggers a new computation and the data itself evolves in response to expert
decisions.
This creates design requirements not present in static analysis tools~--
specifically, state management across evolving data versions, provenance tracking
of the modification sequence, and guidance to navigate the resulting combinatorial
space of computed alternatives.

\subsection{Visual Analytics for Optimization and Planning}
Visual analytics has increasingly been applied to support decision-making tasks involving computational models, simulations, and optimization processes.
In many real-world planning scenarios, optimization algorithms generate candidate solutions that must be analyzed and refined by domain experts.
Visual analytics approaches combine computational analysis with interactive visualization to support a human-in-the-loop exploration of solution spaces~\cite{Keim2010VA,Kohlhammer2012VA}.
The design study methodology has also emphasized the importance of integrating visualization tools with domain workflows to support expert reasoning in complex analytical contexts~\cite{Sedlmair2012DesignStudy}.

Interactive visual steering of simulations and computational models has been
widely explored.
Matković et al.\ introduced visual steering of simulation ensembles, allowing
analysts to interactively guide further computations based on visual
analysis~\cite{Matkovic2015Steering}.
Other work has investigated the interactive exploration of continuous parameter
spaces~\cite{Sedlmair2014ParameterSpace} and the visual analysis of simulation
model outcomes~\cite{TorsneyWeir2011Tuner}.
Coffey et al.\ presented Design by Dragging~\cite{Coffey2013Dragging}, enabling
direct manipulation of simulation inputs and outputs to explore alternative
designs, while World Lines~\cite{Waser2010WorldLines} supports exploration of
alternative simulation futures through interactive steering of scenario parameters.
These approaches operate on \emph{continuous} spaces~-- parameters, geometric
positions, or physical quantities that admit smooth interpolation and
gradient-based navigation.
Our work operates on a \emph{discrete combinatorial} modification space: crossing
operations are atomic, domain-specific actions with no continuous analogue,
which precludes steering and instead motivates the explicit ranking and progressive
disclosure guidance described in Section~\ref{sec:guidance}.
Moreover, unlike steering approaches where the user adjusts inputs to a fixed
simulation, our system uses simulation as an evaluation oracle~-- automatically
triggered after each discrete modification to assess its operational impact.

\subsection{Guidance in Visualization}
Guidance techniques aim to support users in complex analysis tasks by highlighting relevant information, suggesting promising actions, or prioritizing alternatives during the exploration process.
Miksch and colleagues have made foundational contributions to guidance in visual analytics, characterizing guidance as computational assistance that helps users close knowledge gaps during interactive analysis while maintaining human control~\cite{Ceneda2017Guidance}, and surveying guidance approaches to investigate how mixed-initiative interaction between users and systems can support visual data analysis workflows~\cite{Ceneda2019GuidanceSurvey,Stoiber2022OnboardingGuidance}.
Further contributions have explored guidance mechanisms in visual analytics systems, including strategies for integrating guidance into coordinated visual interfaces and supporting users during complex analytical tasks~\cite{May2011Guidance}.
Recent work has proposed frameworks for designing reusable guidance strategies and integrating guidance into interactive visual analytics~\cite{Han2022Guidance,Sperrle2022Lotse,PerezMessina2025Guidance}.
Yang et al.\ proposed an analysis-guided exploration approach for multivariate data, in which computational analysis drives the suggestion of promising next steps during interactive visual exploration~\cite{Yang2007Nugets}.

Our work contributes to this line of research by introducing a three-level guidance framework for exploring aggregated and ranked modification candidates.
In our scenario, potential schedule modifications are grouped by crossing locations, ranked according to a selected evaluation metric, and evaluated using additional contextual attributes that support expert decision-making.
While prior guidance work focuses on supporting general visual analysis tasks, our approach demonstrates how three-level guidance can be applied to iterative modification workflows, where candidate actions must be prioritized and evaluated in the context of evolving system states.
Unlike recommendation systems, which typically suggest a single best action,
guidance in the sense of Ceneda et al.~\cite{Ceneda2017Guidance} preserves human
control by informing rather than prescribing.
Our system embodies this distinction explicitly at Level~3 of the guidance mechanism, where all candidates at a selected location are exposed rather than only the highest-ranked one, allowing experts to override the primary metric ranking based on domain knowledge and secondary operational considerations.
The history mechanism underlying our exploration support relates to work on analytic provenance~\cite{Ragan2016Provenance} and to the analytical reasoning framework of Shrinivasan and van Wijk~\cite{Shrinivasan2008Reasoning}, which introduced a navigation view capturing visualization states as a tree to support revisiting and branching during exploration~-- a principle directly reflected in our provenance tree design.
\section{Background / Railway Scheduling}
\label{sec:background}
This section provides information on the railway planning workflow relevant to our work. 
We first describe the problem of scheduling traction units and the crossing operations, %used to modify the schedules
followed by an overview of the simulation process used to evaluate the robustness of the schedule.
\subsection{Railway Planning and Traction Unit Scheduling}
%%%%%%%%%%%%%%%%%%%%%%%%%%%%%%%%%%%%%%%%%%%%%%%%%%%
%
The optimization problem at hand can be formulated as a Locomotive Scheduling Problem (LSP). 
It involves assigning locomotives~/ traction units to scheduled trains while satisfying operational constraints and minimizing operational costs, which are mainly derived from two factors: the number of traction units used and the distance traveled without hauling trains (deadheading). Deadheading serves to connect disjoint trips, but may also be required for operational reasons, such as local capacity constraints. 

The LSP has been widely studied by both academia and industry. 
Gleaves~\cite{gleaves1957cyclic} proposed one of the earliest algorithms for determining the minimum number of traction units required to maintain a fixed train schedule. Cordeau et al.~\cite{cordeau1998survey} provided one of the earliest surveys on optimization algorithms for train routing and scheduling. A recent survey was published in 2014~\cite{piu2014locomotive}, and Päprer et al.\ recently provided a tutorial that comprehensively synthesizes research in the field~\cite{paprer2025rolling}. Initial research on the LSP focused on exact solutions. Due to the complexity of railway systems, modern applied research also focuses on hybrid approaches, matheuristics, and metaheuristics~\cite{frisch2021solving, habiballahi2022locomotive,allafeepour2023optimization}.
%%%%%%%%%%%%%%%%%%%%%%%%%%%%%%%%%%%%%%%%%%%%%%%%%%%

Railway operations include several interconnected planning tasks such as train, rolling stock, and crew scheduling; here, we focus on traction unit scheduling, which allocates available units to train services. A train can be served by single- or multi-traction, where multi-traction is mostly used for one of two purposes: either to enable faster direction changes because no shunting is required, or to enhance tractive effort especially for heavy trains or steep terrain. 
The most common form of multi-traction is "tandem", where one traction unit is the master, staffed by a driver, while the other unit is remotely controlled from the master. 

Another important aspect of the weekly plan is its maintainability. When a train has a sufficiently long, scheduled waiting time at or near a station with a maintenance facility, a so-called \emph{maintenance window} arises. As trains have to undergo maintenance after specified intervals measured in kilometers driven, it is crucial to plan traction unit schedules in a way that a sufficient number of maintenance windows are available during the opening hours of maintenance facilities.

Frisch et al.~\cite{frisch2021solving} demonstrate that the simultaneous scheduling of locomotives and maintenance windows improves the solution quality, though it increases computational difficulty. % and raises questions about real-world applicability
A further difficulty in a priori maintenance planning is that maintenance resources may be occupied by other types of traction units, tasks, or customers; hence flexibility and adaptability constitute a key component of a successful maintenance planning process.

The optimization considers several operational objectives reflecting the efficiency and robustness of the resulting schedules. These objectives may include factors such as minimizing empty vehicle movements, balancing vehicle utilization, or ensuring operational slack. 
Because these objectives are often conflicting, optimization algorithms can produce a set of Pareto-optimal solutions that represent different trade-offs between the objectives. From this set, domain experts select one candidate solution for further examination and refinement. 
The detailed analysis of this selected schedule forms the starting point of the visual analysis workflow presented in this paper.
\subsection{Railway Simulation}
Simulation models are commonly used to examine how schedules perform under realistic operating conditions.
Unlike optimization models, which typically assume the ideal execution of a timetable, simulation models incorporate disturbances, e.g., delays and operational conflicts.

By simulating the execution of a schedule, planners can identify potential problems such as delay propagation or resource conflicts. 
In the workflow studied in this paper, simulation is applied after schedule generation to assess the robustness of a traction unit schedule and to identify locations where small adjustments to the traction unit assignment may improve operational performance.
The underlying simulation model is a stochastic agent-based model, which was made interactively accessible through the interface. It already existed before the project and is constantly expanded by new use cases \cite{rosslerSimulationOptimizationTraction2020, rosslerAgentbasedModelRobustness2018}.

% Delays & Maintainability
With respect to delay propagation, we distinguish between primary and secondary delays.
Primary delays are delays caused by factors outside the considered system,
such as weather conditions, signal malfunctions, or passenger-related incidents,
whereas secondary delays are propagated delays that arise from, for example, previously delayed traction units or blocked infrastructure.
Since primary delays cannot be avoided, the goal is to minimize secondary delays through robust traction unit schedules.
Accordingly, the number of secondary delays serves as a robustness measure for a traction unit schedule. 
Maintainability is also assessed during the simulation. 
If maintenance windows are not robustly planned, they can shrink due to delays, and maintenance is no longer possible as planned.

% Simulation
To assess the robustness and maintainability of the traction unit schedule, stochastic primary delays are injected at stations and on tracks, and the schedule is simulated. 
The injected delays are sampled from probability distributions fitted to primary delay data extracted from real-world schedule records. 
The underlying disaggregation process separates observed delays into primary and secondary components using network analysis and blocking thresholds~\cite{Schwab2024Disaggregation}, an approach that has also been investigated using graph neural networks~\cite{Viehauser2025GatedGCN}.
The simulation then propagates secondary delays: whenever a train or resource is delayed or a section is blocked, subsequent trains must wait, and the resulting delays are recorded as secondary delays.
The total number of secondary delays is collected for each simulation run. 

% Monte Carlo
For each traction unit schedule, multiple Monte Carlo runs are performed, and the mean number of secondary delays is computed. 
In each run, randomly sampled primary delays are injected. 

% Result
In addition to accumulated secondary delay statistics, the simulation produces a complete circulation plan that captures both the planned and simulated execution of all services, including the resulting delay information for each traction unit.
This output forms the basis for the circulation plan view, which visualizes the planned and simulated schedules side by side, allowing planners to directly identify potential bottlenecks and refinement opportunities.
\section{Analysis Tasks and Design Requirements}
\begin{figure}[t!]
\centering
\includegraphics[width=\linewidth]{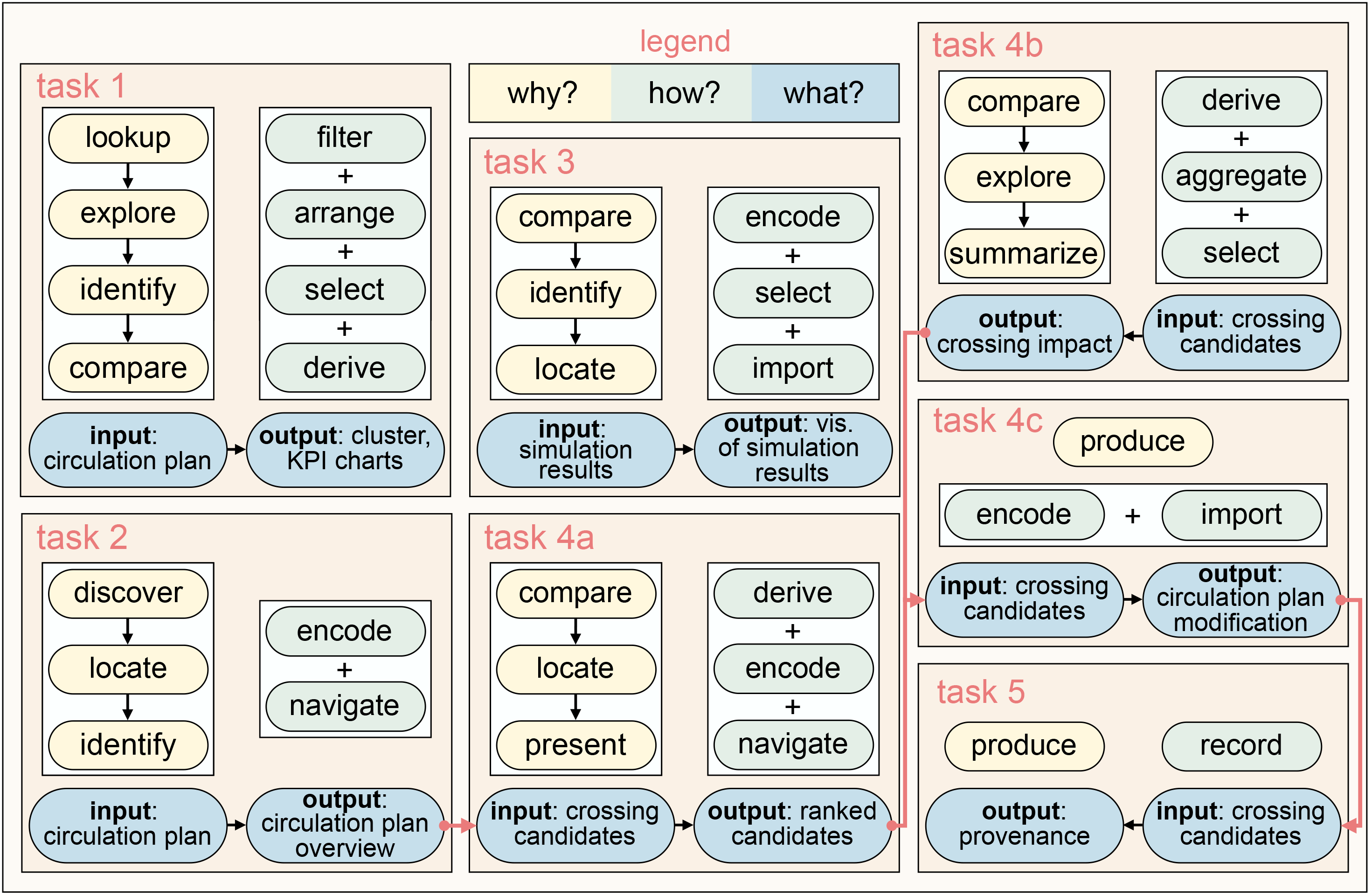}
\caption{Task abstraction for the interactive exploration and refinement of traction unit circulation plans, following the typology of Brehmer and Munzner~\cite{Brehmer2013Typology}. Arrows indicate data flow between tasks. T2 initiates the exploration chain by providing the circulation plan overview; T4a--T4c form a sequential dependency within the crossing workflow, with the resulting plan modification feeding into T5.}
\label{fig:tasks}
\end{figure}

We~-- visualization researchers together with railway scheduling experts~-- jointly derived the following task characterization and design requirements through an iterative process of workflow observations and prototype evaluations. 
Tasks are characterized according to the typology of Brehmer and Munzner~\cite{Brehmer2013Typology}, distinguishing \emph{why}, \emph{what}, and \emph{how}; the formal task abstraction is summarized in Figure~\ref{fig:tasks}, while the descriptions below provide domain-level characterizations. 
The workflow is semi-automatic: computationally intensive steps are performed automatically, while domain experts retain full control over analytical decisions, schedule modifications, and exploration direction. 
%
% -------------------------------------------------------
% TASKS
% -------------------------------------------------------
\subsection*{Task Characterization}
The following five tasks characterize the expert's role in the visual analysis workflow. % \newline
 
\smallskip\hspace*{0.25em}\textbf{T1: understand the structure and operational state of a circulation plan.}
Before any schedule modification can be considered, experts must be able to read and comprehend the circulation plan in the form they are familiar with from daily practice. 
This involves looking up individual traction units and services, exploring the overall structure of assignments across the reference week, analyze load balance, and evaluate the maintainability of the circulation. 
Because the plan encodes complex dependencies, experts also need to inspect per-unit operational parameters to identify problematic units.
%\vspace{0.3\baselineskip} 
\smallskip\par
%and characterize the  schedule state.
\hspace*{0.25em}\textbf{T2: navigate a schedule that exceeds the available display space while preserving global awareness.}
A full seven-day circulation plan for a large cluster of traction units cannot be rendered at full detail on a single screen. 
Experts must be able to navigate to specific regions of the plan~-- whether by time, or by traction unit~-- while retaining a sense of how the currently visible region relates to the overall plan. 
Awareness of the global distribution of operationally relevant indicators must be maintained so that important locations are not overlooked.
%\vspace{0.3\baselineskip} 
\smallskip\par
\hspace*{0.25em}\textbf{T3: assess simulated operational behavior and identify problematic units.}
Because the optimization that produces circulation plans assumes ideal conditions, planned schedules must be evaluated under realistic operating conditions via simulation. 
Experts must be able to compare planned and simulated service execution, identify delay propagation, and locate units whose simulated performance falls below acceptable thresholds, providing the baseline against which any modification is judged.
%\vspace{0.3\baselineskip} 
\smallskip\par
% Experts must be able to compare the planned and simulated execution of services for individual traction units, identify where delays arise and how they propagate, and locate units whose simulated performance falls below acceptable thresholds. This assessment motivates targeted modifications and provides the baseline against which the effect of any change is judged.
%
\hspace*{0.25em}\textbf{T4: identify, evaluate, and apply schedule crossings.}
The primary modification mechanism available to experts is the crossing operation, in which two co-located traction units of the same type exchange their remaining schedules.
%\vspace{0.3\baselineskip} 
\smallskip\par
%Managing crossings involves three interrelated sub-tasks:
\hspace*{1.25em}\textbf{T4a: identify promising crossings.} The expert must be able to survey the full circulation plan and locate stations and time points where crossings are feasible, prioritizing locations where a crossing is estimated to yield the greatest improvement in operational KPIs. This requires aggregating crossing candidates and encoding their estimated impact so that high-potential locations are preattentively salient.
%\vspace{0.3\baselineskip} 
\smallskip\par
\hspace*{1.25em}\textbf{T4b: evaluate individual crossing candidates in detail.} Before committing to a specific crossing, the expert must be able to inspect the expected change in relevant KPIs for each candidate, comparing alternatives and accounting for contextual factors that a summary score alone cannot capture.
%\vspace{0.3\baselineskip} 
\smallskip\par
\hspace*{1.25em}\textbf{T4c: apply a crossing and propagate its consequences.} Once a crossing is selected, the expert must be able to execute it directly in the planning interface. The modified schedule must be recomputed immediately and the updated plan submitted to simulation, so that the operational consequences can be assessed without leaving the analysis.
%\vspace{0.3\baselineskip} 
\smallskip\par
\hspace*{0.25em}\textbf{T5: manage and revisit explored schedule states.}
Applying a crossing produces a new schedule state, and a sequence of crossings constitutes an exploration path through the space of feasible modifications. Because not all paths lead to improvement, experts must be able to track the history of applied crossings, return to earlier states when a modification sequence proves counterproductive, and preserve promising plan variants.
%so that alternative refinement paths can be compared at a later stage.

%
%
% -------------------------------------------------------
% DESIGN REQUIREMENTS
% -------------------------------------------------------
%
\begin{figure*}[t!]
\centering
\includegraphics[width=\textwidth]{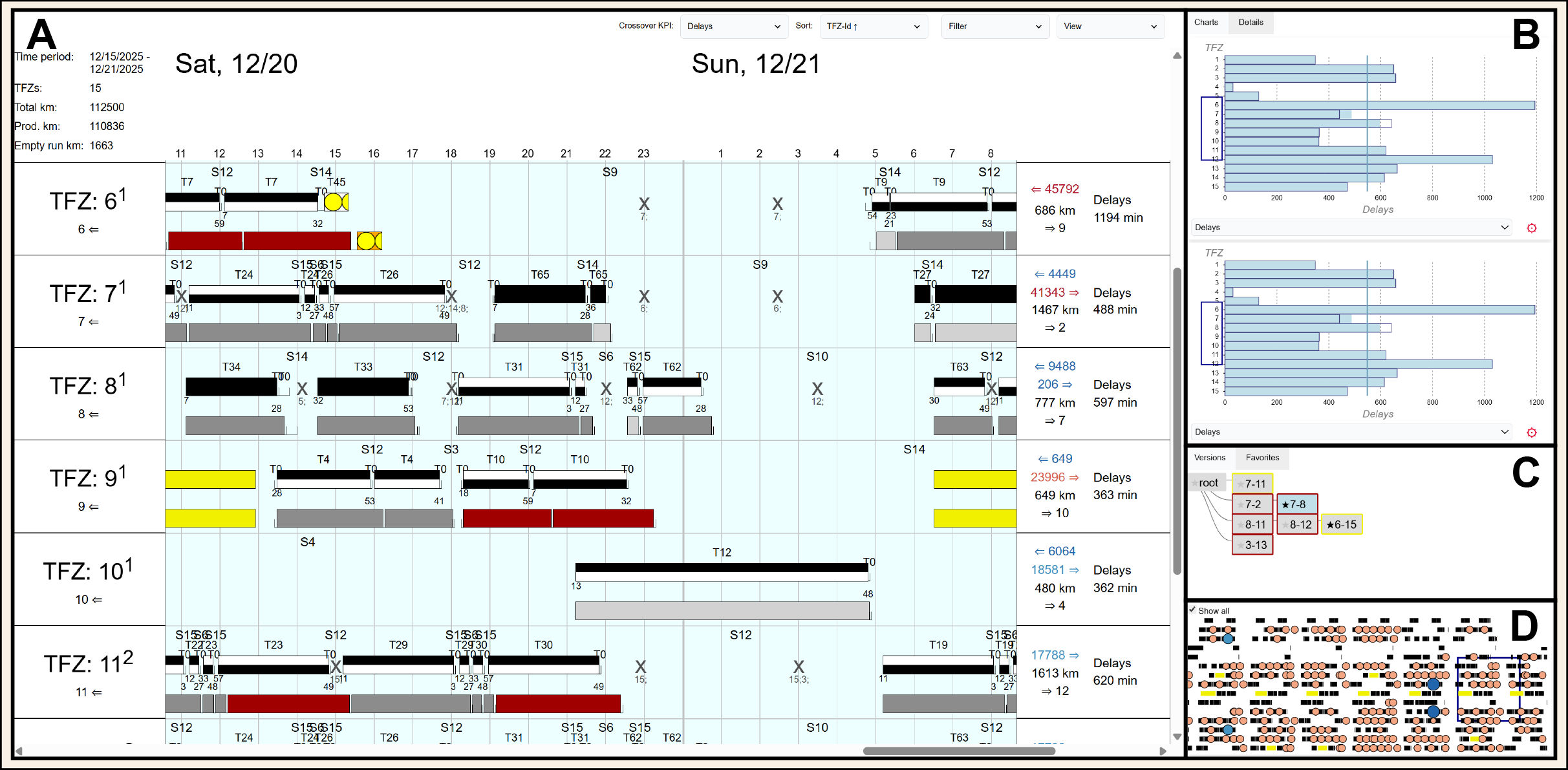}
\caption{Overview of the proposed system, comprising four main panes: 
(A)~Circulation Plan View, which adopts the domain-familiar tabular representation and extends it with interaction, simulation overlay, and crossing indicators; 
(B)~KPI charts, supporting configurable comparison of operational performance indicators across traction units at weekly and per-day granularity; 
(C)~Provenance and Favorites, recording the exploration history as a navigable tree and allowing promising plan variants to be saved and recalled; and
(D)~Circulation Plan Overview, providing a reduced-scale representation of the full plan to support navigation and global spatial awareness.
}
\label{fig:system_overview}
\end{figure*}

\subsection*{Design Requirements}
Based on the tasks, we identified the following design requirements.%\vspace{1em} 
\smallskip\par
\hspace{0.25em}\textbf{R1: present the circulation plan in its domain-familiar form and support flexible exploration of its contents} \emph{(T1)}.
The plan must be rendered in the tabular format that planning experts already use, with one traction unit per row and time encoded horizontally. To manage the complexity of a large fleet, the system must support sorting and filtering of rows by user-defined criteria, toggling between compact and full row modes, and pinning selected rows during sorting and scrolling.%
%\vspace{0.3\baselineskip} 
\smallskip\par
%Hover and selection interactions must expose additional detail on demand.
\hspace{0.25em}\textbf{R2: enable the comparison of per-unit operational profiles across the fleet at multiple temporal granularities, spatially aligned with the plan} \emph{(T1)}.
Experts need to compare operational parameters across all traction units simultaneously. 
This comparison must be available at two temporal granularities: an aggregated view covering the full reference week, supporting overall fleet-level comparison, and a day-level view in which per-day values are spatially aligned across units.
%, supporting both day-specific comparison across units and inspection of load distribution within a unit's week. 
Profile displays must remain synchronized with the row order and scroll position.%\vspace{0.3\baselineskip} 
\smallskip\par
\hspace{0.25em}\textbf{R3: support the navigation of the full plan while maintaining awareness of the global layout and the distribution of operationally relevant indicators} \emph{(T2)}.
The system must provide a representation of the complete circulation plan at a scale at which it fits in the display, allowing experts to perceive the global structure of the schedule. 
This overview must indicate the part of the plan currently visible in the main view, and  encode the positions and estimated impact of crossings so that high-potential regions are visible regardless of the current viewport.% \vspace{0.3\baselineskip} 
\smallskip\par
%Navigation to a target region must be possible directly from the overview.
\hspace{0.25em}\textbf{R4: overlay a simulated execution on the planned schedule to expose delay propagation} \emph{(T3)}.
For each plan state, the system must render the simulated service execution alongside the planned schedule within the same view, using visual differentiation to distinguish planned from simulated timings. The representation must make delay propagation legible at the level of individual traction units, enabling experts to relate operational problems to specific services and time windows.%\vspace{0.3\baselineskip} 
\smallskip\par
\hspace{0.25em}\textbf{R5: provide a three-level visual guidance mechanism for identifying and evaluating crossing candidates} \emph{(T4a, T4b)}.
Crossing candidates must be surfaced at two levels of detail. At the overview level, candidate locations must be aggregated and their estimated KPI impact encoded visually, allowing experts to direct attention to the most promising regions of the plan without examining each candidate individually. 
At the detail level, the system must present the specific KPI changes associated with individual candidates at a selected location, supporting informed comparison before a decision is made.%\vspace{0.3\baselineskip} 
\smallskip\par
%The guidance must remain coherent as the plan evolves through successive modifications.
\hspace{0.25em}\textbf{R6: support the direct application of crossings in the interface and automatic propagation of consequences} \emph{(T4c)}. 
Experts must be able to apply a selected crossing without leaving the analysis interface. Upon application, the system must immediately recompute the affected schedule and automatically initiate a new simulation, updating all views to reflect the modified plan state so that the effect of the crossing can be assessed in context.%\vspace{0.3\baselineskip} 
\smallskip\par
\hspace{0.25em}\textbf{R7: maintain a structured history of explored states and support non-linear exploration of modification paths} \emph{(T5)}. 
The system must record each applied crossing as a discrete step in an exploration history, allowing experts to review the sequence of modifications leading to the current state. It must be possible to revert to any prior state in the history. 
Additionally, experts must be able to designate specific plan variants as favorites and retrieve them subsequently, to compare alternative refinement paths.
%across separate exploration sessions.

% 
%

%

%
%
\section{Visual and Interaction Design}
\label{sec:design}
This section describes the visual mappings and interaction design of the proposed approach.
The system is organized into four main panes, as shown in
Figure~\ref{fig:system_overview}.
The {Circulation Plan View} (Section~\ref{sec:circ_plan}) is the central element, presenting scheduled and simulated services in the domain-familiar tabular form.
The {KPI charts} (Section~\ref{sec:kpi_charts}) provide spatially aligned performance profiles per unit to support fleet-level comparison.
The {Circulation Plan Overview} (Section~\ref{sec:overview}) supports navigation and global spatial awareness throughout the reference week.
The {Provenance and Favorites} panel (Section~\ref{sec:provenance}) tracks the exploration history and enables saving and recalling of plan variants.
The three-level guidance mechanism, which spans multiple panes and constitutes a distinct contribution, is described separately in Section~\ref{sec:guidance}.
\subsection{Circulation Plan View}
\label{sec:circ_plan}
The circulation plan view forms the central element of the interface and is depicted in Figure~\ref{fig:system_overview}A.
It shows the circulation plan in its standard tabular form, with traction units arranged in rows and their assigned train schedules visualized horizontally over the course of one reference week \emph{(R1)}. 
The tabular circulation plan format is a domain standard adopted directly from existing planning practice; a schematic illustration is provided in Appendix~\ref{suppl:prevworkflow}.
A summary panel in the upper left displays the reference week, number of traction units, and relevant operational KPIs including overall production kilometers and deadheading kilometers, providing immediate context for the plan under analysis.

\begin{figure}[t!]
\centering
\includegraphics[width=\linewidth]{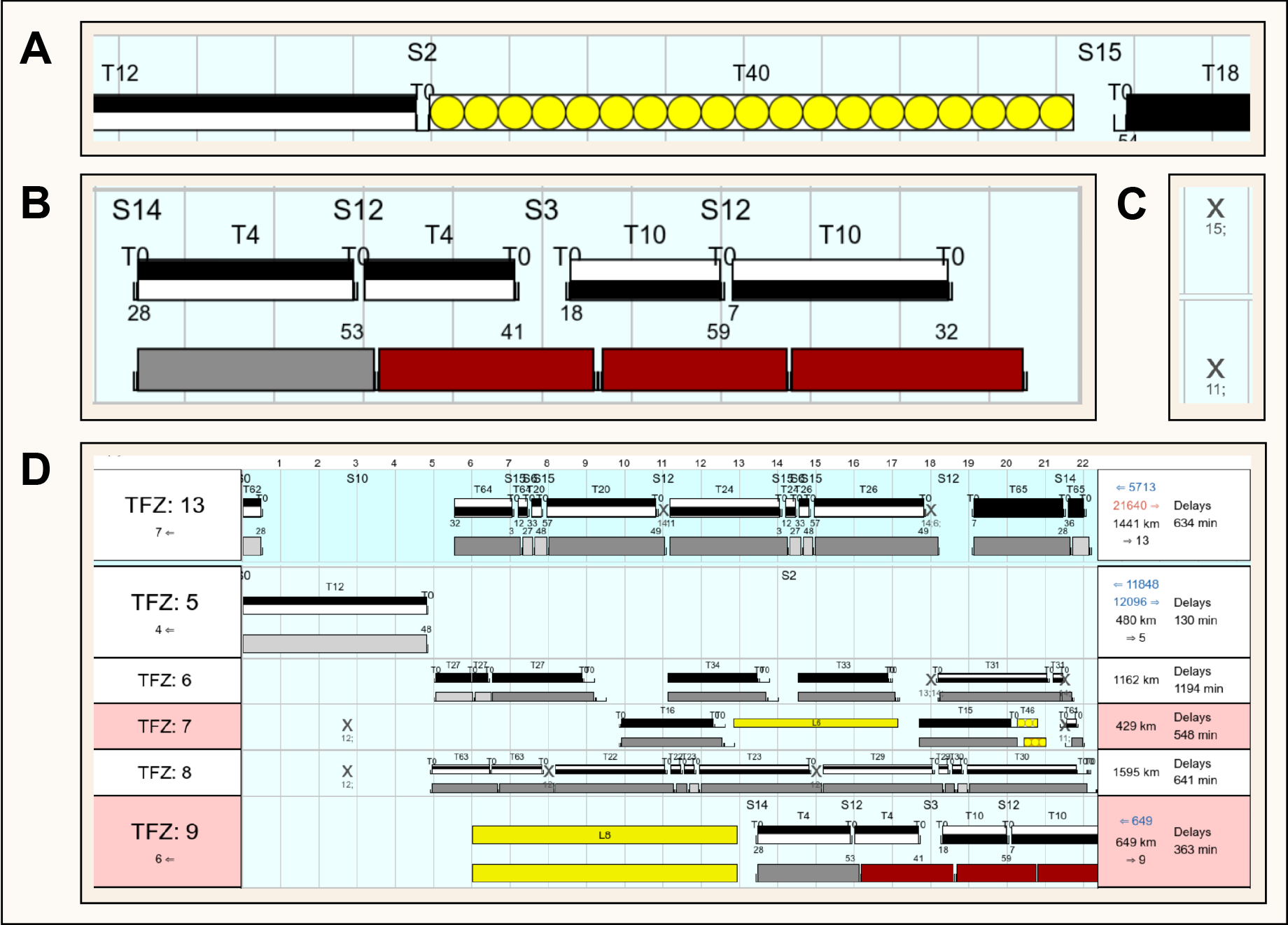}
\caption{Details of the Circulation Plan View: 
(A)~Example of consecutive runs of one traction unit, showing traction operation in a tandem operation (black and white bar), a deadheading (empty) run (bar with yellow circles), and a subsequent single-unit operation (black bar);
(B)~Train run bars above simulated data, color-coded according to the magnitude of the delay from gray to red; 
(C)~Crossing possibilities for traction units 11 and 15;  
(D)~Comparison of compact (traction units 6-8) and full row representations, a pinned row (traction unit 13) and two highlighted rows.}
\label{fig:zoom}
\end{figure}

Each traction unit occupies a single row, with individual services~-- referred to as train runs~-- represented as horizontal bars spanning their respective duration (Figure~\ref{fig:zoom}A).
The bars are color-coded according to the mode of traction (e.g., tandem, single-traction).
Deadheading, in which a traction unit travels without providing service, are highlighted using yellow circles, and possible maintenance windows are rendered as yellow bars, indicating time intervals in which a traction unit is located near a maintenance facility with sufficient dwell time.
The plan can be explored by scrolling across both traction units and time. Circulation plans are generated in a weekly cyclic structure, requiring traction units within a cluster to collectively return to their starting stations at the end of the reference week, though not necessarily as the same physical unit (\emph{R1}).
The traction unit assigned in the preceding week is also listed for each row.

For direct comparison of planned and simulated behavior, simulated train runs~-- generated by stochastically introducing delays based on historical data~-- are rendered as a second bar directly beneath the scheduled train runs, see Figure~\ref{fig:zoom}B (\emph{R4}).
Simulation runs are color-coded from gray to red according to the magnitude of the accumulated delay, with light gray indicating delays below 10~minutes, dark gray indicating delays below 30~minutes and red indicating any longer delay, making delay propagation immediately legible at the level of individual services and traction units. This categorization follows commonly used threshold values. Depending on the operational context (e.g., passenger or freight services), different thresholds may be applied.

Since traction units within a cluster are always of the same type, any two units co-located at the same station at the same time can exchange their remaining schedules via a crossing operation (Figure~\ref{fig:zoom}C).
Potential crossings are indicated directly in the plan by a gray `x', annotated with the identifiers of the traction units involved, allowing experts to locate candidate modification points at a glance (\emph{R5}).
The visual ranking and aggregation of these candidates is described in detail in Section~\ref{sec:guidance}.

A panel to the right of each row displays the per-day distances to the next and from the previous maintenance window, as well as the total predicted delay accumulated over the reference week, providing a concise operational summary per traction unit (\emph{R2}).

Hovering over train runs, simulation runs, crossings, and maintenance windows reveals detailed information on demand (\emph{R1}).
Rows can be pinned to remain visible at the top of the plan regardless of the current sort order, highlighted for cross-row comparison, hidden to reduce visual complexity, or reduced in size using a compact mode (Figure~\ref{fig:zoom}D).
Filtering and sorting by user-defined criteria~-- including delay magnitude, and distance to maintenance~-- allow experts to restructure the plan to surface traction units of interest, with all linked views updating accordingly (\emph{R1}).
Additional display options allow labels, simulation data, or specific traction units to be toggled on or off to manage the complexity of the view.
Train runs are labeled with station names and train numbers, and traction units are identified by their assigned identifiers in the row headers.
\begin{figure}[t!]
\centering
\includegraphics[width=\linewidth]{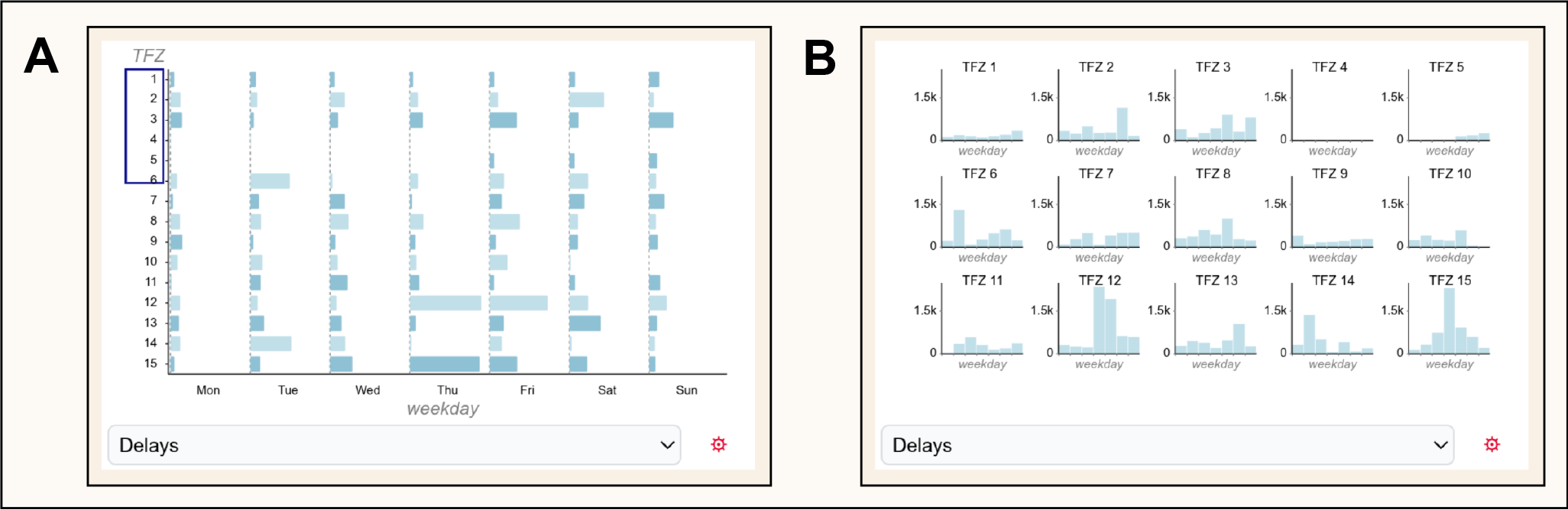}
\caption{KPI chart configurations, shown here for predicted delays:
(A)~Per-day mode, in which each bar is split into seven segments aligned by weekday, supporting both day-level comparison across units and inspection of load distribution in a unit's week. 
(B)~Small-multiples mode, providing a dedicated miniature chart per traction unit with weekdays on the x-axis for detailed temporal analysis. 
%In (A), (B), and (C), a blue rectangle on the corresponding axis indicates the subset of units currently visible in the circulation plan.
}
\label{fig:charts}
\end{figure}

\subsection{KPI Charts}
\label{sec:kpi_charts}
To enable a comparison of operational KPIs between traction units at multiple levels of temporal granularity, two configurable charts are integrated into a side panel adjacent to the circulation plan, shown in Figure~\ref{fig:system_overview}B and
Figure~\ref{fig:charts} (\emph{R2}).
For each chart, the displayed parameter can be selected from a range of operational indicators, including production kilometers, delay minutes, total run kilometers, and distance to the next maintenance window.
A settings panel allows users to overlay an average reference line and adjust the scale to accommodate larger fleets for structured comparison.
 
The charts' layout evolved in an iterative design process.
An initial vertical bar chart, with traction units on the x-axis and KPI values on the y-axis, was replaced by a horizontal representation
(Figure~\ref{fig:system_overview}B) in which traction units are arranged on the vertical axis, directly mirroring the row structure of the circulation plan (\emph{R2}).
This alignment ensures that the KPI profile of each traction unit is spatially co-located with its row in the plan, enabling direct visual comparison without mental re-mapping between views.
 
In addition to aggregated weekly values, we support two further display modes to address the dual temporal granularity required by \emph{R2}.
The per-day mode (Figure~\ref{fig:charts}A) divides each bar into seven segments aligned by weekday, allowing experts to compare the same day between traction units or inspect the load distribution within a single unit's week.
The small-multiples mode (Figure~\ref{fig:charts}B) displays a separate miniature chart per traction unit, with weekdays on the x-axis and the selected KPI on the y-axis, supporting a detailed inspection of temporal patterns for individual units.
 
All chart elements are synchronized with the circulation plan: when the plan is sorted or filtered, the chart rows reorder accordingly, and a blue rectangle on the vertical axis marks the subset of traction units currently visible in the main view
(\emph{R1}).
A detail panel, accessible through a dedicated tab, provides an extended list of operational KPIs for both the planned schedule and the simulated data, supporting deeper per-unit analysis beyond what the charts display at a glance (\emph{R2}).

\subsection{Provenance and Favorites}
\label{sec:provenance}
Applying a crossing produces a new schedule state, and a sequence of crossings constitutes an exploration path through the space of feasible modifications.
To support a structured, non-linear exploration of such paths, the system maintains a provenance tree that records every schedule state visited during a session (\emph{R7}), see Figure~\ref{fig:system_overview}C.

Each node in the tree represents a distinct schedule state, labeled with the identifiers of the two traction units that were crossed.
The root node represents the original unmodified plan.
Edges connect parent states to the child states derived from them by applying a crossing.
The currently active state is highlighted, and all views~-- the circulation plan, the KPI charts, and the overview~-- reflect the schedule corresponding to that state.
To support orientation within the tree, each node is framed with a color reflecting the KPI objective under which it was produced: red frames indicate nodes created while optimizing for delays, consistent with the red color used to render simulated delays in the circulation plan; yellow frames indicate nodes created while optimizing for maintenance, consistent with the yellow used to render maintenance windows.

Branching occurs naturally when the user selects any existing node as the current state and applies a new crossing: rather than overwriting the previous exploration path, the system creates a new branch from the selected node.
This allows experts to revisit earlier states when a sequence of modifications proves unproductive and explore alternative refinement paths without losing previously computed states (\emph{R7}).

To support the direct comparison between schedule states, the user can select any second node in the provenance tree as a reference state.
The KPI values of the reference state are then overlaid on the KPI charts as outlined rectangles rendered on top of the bars of the currently active state (\emph{R7}), see Figure~\ref{fig:system_overview}B.
This allows experts to immediately perceive which traction units improved or degraded the reference state, supporting informed decisions about whether to continue along the current exploration path or revert to an earlier one.

Schedule states that are found to be particularly promising can be marked as favorites using a star indicator on the corresponding node.
Favorite states are collected in a dedicated \emph{Favorites} tab, presented as a list in which each entry shows the state identifier alongside a short user-provided description (\emph{R7}).
This description field allows experts to record the analytical reason for saving a state~-- for example, noting that a particular configuration eliminates delays at a specific station~-- making it possible to recall and compare promising variants across longer exploration sessions.
Selecting any favorite or any node immediately restores the corresponding schedule state.

\subsection{Circulation Plan Overview}
\label{sec:overview}
Traction unit clusters can comprise a large number of units whose full seven-day schedule cannot be rendered in full detail on the available display area. 
To support navigation and maintain global spatial awareness, a reduced-scale overview of the complete circulation plan is integrated into the side panel (\emph{R3}), shown in Figure~\ref{fig:system_overview}D.
 
The overview renders a simplified representation of all train runs across the reference week, with maintenance windows highlighted in yellow to preserve operationally relevant landmarks.
A blue rectangle overlaid on the overview indicates the portion of the plan currently visible in the main view; users can reposition this rectangle by direct interaction to pan the main view to any region of the plan (\emph{R3}).
 
In addition to supporting navigation, the overview encodes the distribution of crossing candidates across the plan by rendering crossing locations as colored circles of varying size (\emph{R5}).
Since each crossing involves exactly two traction units, every crossing is represented by a vertically aligned pair of circles~-- one in each of the two traction unit rows involved.
Both circles in a pair share the same horizontal position, size, and color, as these properties are derived from the KPI impact of the crossing itself rather than from either individual unit.
The vertical alignment implicitly groups the two circles as a single crossing without requiring additional visual connectors.
Blue circles indicate locations where a crossing is estimated to improve the selected KPI, while orange circles indicate an estimated negative impact; circle size encodes the magnitude of the estimated effect.
This allows experts to identify promising modification regions at a glance without having to scroll through the full plan. 
The encoding and ranking logic underlying these indicators is described in detail in Section~\ref{sec:guidance}.
\subsection{Three-level Guidance}
\label{sec:guidance}
A key challenge in exploring crossing modifications is that the number of possible crossing sequences grows exponentially with the number of steps, making an exhaustive evaluation infeasible in practice.
Our guidance mechanism therefore operates one step ahead: for each crossing candidate visible in the current schedule state, the effect of applying that crossing is simulated independently, and the resulting KPI change is used to rank and visualize the candidates.
Crossings that would become available after applying a modification are not pre-computed, as evaluating all possible sequences to arbitrary depth would not be feasible for real-world network sizes.
This design grounds the guidance in the current state and updates it automatically whenever a new state is reached, keeping the computational cost bounded while providing actionable, locally optimal recommendations at each step (\emph{R5, R6}).

The three-level structure of the guidance mechanism is designed to be generalizable beyond the railway scheduling domain.
At the first level, modification candidates are aggregated by a grouping criterion and encoded by a primary ranking metric, allowing high-potential groups to be visually distinct and quickly
identifiable across the full problem space.
At the second level, all candidates within a selected group are exposed and ranked by the same metric, supporting local comparison without commitment.
At the third level, a full per-candidate breakdown is provided, including secondary attributes that the primary metric alone cannot capture and that may ultimately govern the expert's decision.
This staged disclosure pattern is likely applicable to similar semi-automatic, human-in-the-loop workflows in which candidate actions can be grouped, ranked by a primary metric, and evaluated using additional contextual information before committing to a modification.

The guidance operates across the Circulation Plan Overview and the Circulation Plan View, progressively exposing more information as the user focuses attention on a region of interest.
\paragraph{Level 1: spatial overview of crossing candidates.}
In our case, the grouping criterion is the crossing location: crossing candidates are aggregated by station and time point; each group is represented in the Circulation Plan Overview as a colored circle of varying size (\emph{R5}).
For each location, the candidate with the greatest estimated KPI improvement is identified, and the circle encoding reflects this best-case estimation.
Circle size and color shade together encode the magnitude of the estimated KPI impact: larger and darker circles indicate a greater estimated effect, while smaller and lighter circles indicate a lesser one.
Color hue encodes the direction: blue indicates locations where at least one crossing is estimated to improve the selected KPI, and orange indicates locations with an estimated negative impact.
Colors were selected from the ColorBrewer diverging palette to ensure distinguishability for users with a color vision deficiency~\cite{Harrower2003ColorBrewer}.
Since in practice the number of crossing locations can be large and experts are primarily interested in locations where improvements are possible, a filter option allows users to display only blue circles, reducing visual clutter and focusing attention on actionable candidates.
As illustrated in Figure~\ref{fig:system_overview}D, the unfiltered overview reveals a large number of crossing locations.
Without the guidance mechanism, experts would need to inspect each location individually to assess its potential; the color and size encoding reduces this to a pre-attentive scan, and the filter option further reduces visual clutter by hiding locations with a negative estimated impact.
In the remaining use cases and figures, the filter is enabled, displaying only blue circles.
%This encoding makes high-potential modification regions immediately salient across the full reference week, allowing experts to direct attention to promising locations without scrolling through the plan.
It is important to note that the circle reflects only the best available candidate at each location.
%--- the number of candidates and the distribution of their impacts are not conveyed at this level.
%
\paragraph{Level 2: candidate ranking in the Circulation Plan View.}
In a selected location, all crossing candidates are exposed in the Circulation Plan View as traction unit identifiers rendered at the crossing point (\emph{R5}).
The font size of each identifier encodes the estimated KPI impact of the corresponding crossing by the primary ranking metric: larger font indicates a more favorable outcome, smaller font a less favorable or negative one.
This allows experts to compare all available candidates at a location at a glance and identify the most promising one without yet committing to a modification.
Where multiple traction units are available for crossing, all are shown simultaneously, giving a complete picture of the options at that point in the plan.
\paragraph{Level 3: detailed per-candidate evaluation.}
Clicking on a crossing location reveals the full per-candidate breakdown for \emph{all} candidates at that location, not only the highest-ranked one (\emph{R6}).
This is a deliberate design decision: while Level 2 uses font size to direct attention to the most promising candidate, Level 3 exposes the complete picture so that experts can make an informed choice based on secondary attributes.
% that the primary ranking metric alone cannot capture.
The breakdown is provided per traction unit involved in the crossing: the change in delay for each unit is shown individually, with the aggregated net gain across both units.
This decomposition exposes secondary attributes that the primary ranking metric alone cannot capture~-- for example, a crossing where one unit gains ten minutes and the other loses ten minutes yields a net gain of zero, yet may still be preferred if the improvement is operationally more significant than the degradation, or if secondary indicators favor it.
Additional operational KPIs beyond the primary ranking metric are also shown for each candidate, allowing experts to evaluate options along multiple dimensions before committing (\emph{R6}).
Based on this information, the expert selects the crossing to execute, upon which the schedule is recomputed and a new simulation is automatically triggered to evaluate the updated state (\emph{R7}).
If none of the candidates are deemed sufficiently promising, the expert can dismiss the detail view and consult other locations.
\begin{figure}[t!]
\centering
\includegraphics[width=\linewidth]{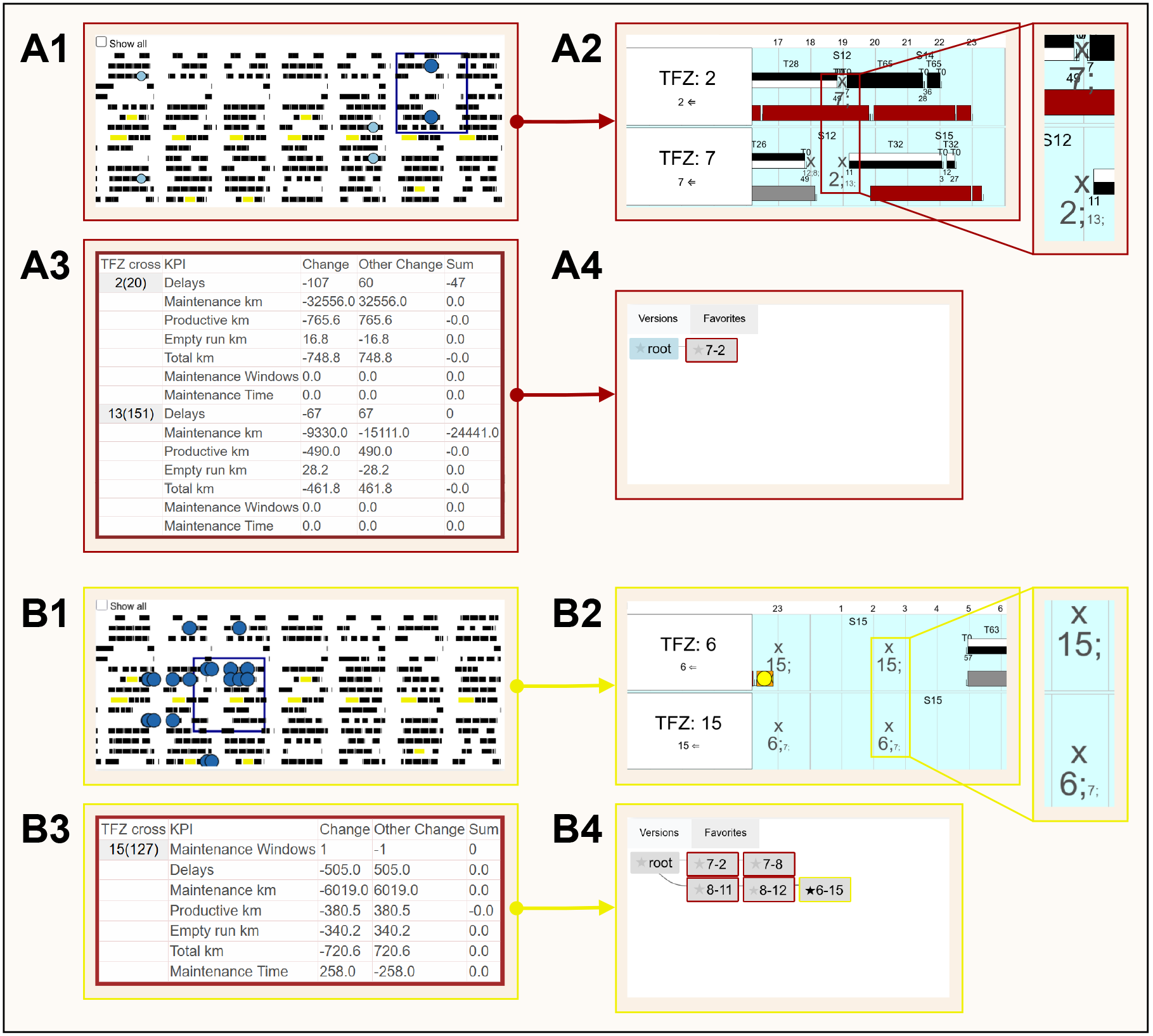}
\caption{
Three-level guidance steps for Use Case~1 (A, red frames) and Use Case~3 (B, yellow frames), with frame colors reflecting the active KPI objective: Delays (red) and Maintenance Windows (yellow). (A1,~B1)~Circulation Plan Overview with blue circles at crossing locations; the blue focus rectangle indicates the region currently visible in the Circulation Plan View (not shown in this figure). (A2,~B2)~Zoomed Circulation Plan View at the selected crossing location. (A3,~B3)~Detailed view showing the exact changes of different KPIs for each crossing candidate. (A4,~B4)~Provenance tree after executing the crossing.
}

\label{fig:usecases}
\end{figure}
\section{Use Cases}
\label{sec:usecase}
We present three cases, demonstrating how the system supports the iterative exploration and refinement of a traction unit circulation plan with respect to two operational KPIs: minimizing predicted delays and optimizing the distribution of maintenance opportunities in the reference week.
All cases are demonstrated in the accompanying video.
%
% -------------------------------------------------------
%
\subsection{Data}
\label{sec:data}
The circulation plan under consideration is based on a schedule provided by a major Austrian railway operator.
It consists of a cluster of 15 traction units with schedules spanning over one reference week from Dec.\ 15, 2025, to Dec.\ 21.
The plan represents an ideal-case schedule and serves as the starting point for the visual analysis.

For each schedule state, stochastic primary delays are injected, and the plan is simulated using the approach described in Section~\ref{sec:background}.
The injected delays are derived from historical delay data spanning six months of real-world operations.
The simulation output~-- including both the planned and simulated service execution~-- forms the data basis for the Circulation Plan View.
Whenever a crossing is applied, a new schedule state is produced and submitted to simulation automatically, which yields a new node in the provenance tree.
All states are retained, allowing experts to freely navigate between them and compare any two states using the reference overlay in the KPI charts.

For confidentiality reasons, train numbers and station names have been anonymized in all figures.
%
% -------------------------------------------------------
% -------------------------------------------------------
% 
\subsection{Use Case~1: Minimizing Predicted Delays}
\label{sec:uc1}
The aim of the first use case is to improve the circulation plan with respect to predicted delays using the three-level guidance system.
The planners select \emph{Delays} as the target KPI and enable the filter to display only crossings with a positive estimated impact, reducing visual clutter in the overview.
 
\paragraph{Level 1: identifying promising locations.}
Using the Circulation Plan Overview, the planners observe three crossing locations, represented by vertically aligned pairs of blue circles, see Figure~\ref{fig:usecases}A1.
The largest and darkest pair indicates the crossing with the best estimated delay reduction.
The planners reposition the blue focus rectangle in the overview to adjust the Circulation Plan View to this location.
 
\paragraph{Level 2: comparing candidates at the selected location.}
At the selected crossing location, two candidates are visible in the Circulation Plan View: a crossing between traction units 2 and 7, and an alternative crossing between units 7 and 13 (Figure~\ref{fig:usecases}A2).
The larger font size of the identifiers for units 2 and 7 indicates that this crossing has a higher estimated impact on the delay KPI.
The planners therefore focus their attention on the crossing between units 2 and 7, which occurs on Saturday at 7\,pm.
 
\paragraph{Level 3: evaluating the selected crossing.}
Clicking on the crossing location opens the detailed view, which displays exact KPI changes for all crossing candidates at that location~-- in this case, both crossings between units 2 and 7 and between 7 and 13 (Figure~\ref{fig:usecases}A3).
This allows the planners to compare the two options across all available KPIs, not only the primary delay metric used for ranking at Levels 1 and 2.
The view presents the induced change for each traction unit individually, as well as the aggregated net effect across both units.
The rows include the delay-related KPI followed by additional operational KPIs such as maintenance kilometers, productive kilometers, deadheading kilometers, total run kilometers, maintenance windows, and maintenance time.
Confirming that the crossing between units 2 and 7 offers a better outcome across the relevant KPIs, it is executed.
\paragraph{Outcome.} After executing the crossing, the circulation plan is updated, a new simulation is automatically triggered, and the KPI charts are refreshed to reflect the modified state.
The provenance tree records the new state as a node labeled with the identifiers of the crossed units (Figure~\ref{fig:usecases}A4).
To maintain visual consistency with the Circulation Plan View, the node is framed with a red rectangle, reflecting that it was produced under the delay KPI objective~-- the same color used to render simulated delays in the plan.
The planners inspect the updated KPI charts to confirm the improvement and save the state as a favorite for later comparison.
%------------------------------------------------------
%
\subsection{Use Case~2: Optimizing Tandem Continuity}
\label{sec:uc2}
The second case shows how we support a modification motivated by operational efficiency rather than a direct KPI improvement.

\paragraph{Identifying the modification opportunity.}
Building on the state produced in Use Case~1, the planners identify an opportunity to improve tandem continuity involving traction units 7 and 8 on Saturday at 6\,pm. As shown in Figure~\ref{fig:system_overview}A, traction unit~7 operates as the master unit in a tandem on train T26 from station S15 to S12, indicated by the split black and white bar in its row. Unit~8 subsequently operates as the master in a tandem on train T31 on the return trip from S12 to S15. Rather than exchanging the master unit between the two trains, a crossing at this location allows unit~7 to continue on the return trip as the remotely controlled unit on train T31, eliminating an unnecessary locomotive exchange and preserving the existing tandem pairing.

\paragraph{Evaluating the trade-off.}
The planners open the detail view for the crossing between units 7 and 8.
The delay KPI indicates a slight increase of 3 minutes for the entire traction unit cluster as a net effect of the crossing.
Despite this marginal degradation of the delay KPI, the planners judge the improvement in tandem continuity to be operationally preferable, eliminating an unnecessary locomotive exchange.

\paragraph{Outcome.} The planners execute the crossing and the system produces a new schedule state.
The provenance tree now contains two nodes branching from the root: the delay-optimized state from Use Case~1 and the tandem-optimized state from Use Case~2.
Using the reference overlay in the KPI charts, the planners compare the two states directly, confirming that the tandem crossing introduces only a marginal delay penalty while achieving the intended operational benefit.

%
% -------------------------------------------------------
%
\subsection{Use Case~3: Balancing Maintenance Windows}
\label{sec:uc3}
The third case starts with the initial circulation plan, representing a separate exploration branch in the provenance tree, independent of the modifications in Use Cases~1 and~2.
The goal is to optimize the distribution of maintenance window opportunities across all traction units within the reference week.
To establish a suitable starting point, planners first select \emph{Delays} as the target KPI and perform two delay-reducing crossings.
The first crossing, between units 8 and 11 on Friday at 7\,pm, corresponds to the second-largest delay reduction (16~minutes), as indicated by the second-largest and second-darkest pair of blue circles in the Circulation Plan Overview; it also simultaneously resolves a tandem continuity issue.
In the resulting state, a second favorable crossing for delay reduction emerges between units 8 and 12 at the same location, which is also executed.
Thereafter, no further delay-reducing crossings are identified, and planners shift their focus to maintenance window balancing by selecting \emph{Maintenance Window} as the target KPI.
\paragraph{Levels 1 \& 2: identifying and comparing promising crossing candidates.}
By activating the filter to display only crossings that improve maintenance window balance, seven candidate crossings are displayed in the Circulation Plan Overview, as shown in Figure~\ref{fig:usecases}B1.
After visually inspecting these candidates, the planners select a crossing between traction units 6 and 15 on Wednesday at 2:30,am (Figure~\ref{fig:usecases}B2).
\paragraph{Level 3: evaluating the selected crossing.}
The detail view reveals that the execution of this crossing gives traction unit~6, originally without a maintenance window in the reference week, one maintenance possibility on Wednesday between 12:51 and 5:09\,pm, while traction unit~15, which initially had two, loses one (Figure~\ref{fig:usecases}B3).
This results in a more balanced distribution of maintenance opportunities, while all other operational KPIs are unaffected.
The crossing is executed.
\paragraph{Outcome.} The execution of the crossing updates the circulation plan, triggers a new simulation, and refreshes the KPI charts.
The provenance tree now contains three additional nodes: two framed in red (delay KPI) and one framed in yellow (maintenance window), consistent with the colors used in the circulation plan view (Figure~\ref{fig:usecases}B4).

% %
% -------------------------------------------------------
% 
\subsection{Expert Assessment}
\label{sec:uc_summary}
The three use cases illustrate the most important system operations: guided identification of delay-reducing crossings,  facilitated by three-level guidance, expert-driven modifications, motivated by operational considerations beyond the primary KPI, and iterative exploration of a separate refinement path, targeting a different objective.
Together, they demonstrate how the provenance tree supports non-linear exploration and how the reference overlay enables direct comparison between alternative schedule states.
 
From our experience as planning practitioners, the capability shown in Use Case~1 substantially improves the current approach of manually comparing crossing candidates.
The Circulation Plan Overview, combined with the color-coded circle encoding of the guidance mechanism, provides an intuitive and efficient way to identify beneficial modifications across the whole reference week~-- a task that previously required tedious manual inspection~-- while the provenance tree enables the easy comparison between alternative exploration paths.
 
The modification demonstrated in Use Case~2 reflects a typical procedure that arises regularly in practice, particularly at early stages of scheduling. While existing planning systems provide some technological support for this task, the explicit highlighting of crossing candidates directly within the plan, combined with the detailed view, represents a novel approach that accelerates the identification and evaluation of tandem continuity improvements.
A precise time estimate for the improvement over the current workflow is difficult to provide, as the time required for manual comparisons strongly depends on the problem size due to the combinatorial complexity of the crossing operations. 
Nevertheless, we find the time savings to be significant, as the guidance mechanism eliminates the need to manually inspect and compare crossing candidates one by one.
 
Maintenance window balancing, as demonstrated in Use Case~3, gains particular importance when combined with the capabilities of Use Cases~1 and~2.
Executing a crossing to reduce delays could result in an unmaintainable schedule by shifting all maintenance opportunities to a single traction unit; the follow-up operation in Use Case~3 addresses this issue efficiently.
The provenance tree is particularly valuable in this context, as it provides a direct causal link between modifications, supporting structured reasoning about their consequences.
\section{Discussion}
\label{sec:discussion}
We reflect on the scope and limitations of the presented evaluation, the design decisions underlying the guidance mechanism, and the potential generalization of the proposed approach beyond the railway domain.
\paragraph{Evaluation scope and planned validation.}
The presented use cases are grounded in real operational data and reflect the conditions of actual planning practice. However, the evaluation is necessarily limited in scope: the observations reflect the experience of two experts working with one specific traction unit cluster, and do not constitute a formal user study with independent participants. A broader evaluation involving additional planning experts is planned before integrating the system into the operational workflow. Such integration requires careful alignment with existing tooling, training, and organizational processes~-- a transition that must be managed carefully, as is common in applied visualization research in safety-critical operational domains. Following design study methodology~\cite{Sedlmair2012DesignStudy}, we consider this a necessary first step toward broader empirical validation.
\paragraph{One-step-ahead guidance and local optimality.}
The three-level guidance mechanism operates one step ahead: it evaluates the immediate KPI impact of each crossing candidate in the current state but does not pre-compute the consequences of subsequent modifications.
This design keeps the computational cost bounded and ensures that the guidance is always grounded in the current state of the plan, but it means that the recommended candidates are locally optimal, rather than globally.
In the presented cases, satisfactory outcomes were reached within two to three crossing steps per exploration branch, suggesting that the local horizon is sufficient for typical problem sizes; the provenance tree further mitigates the limitation by making backtracking easy.
Extending the guidance using beam search~\cite{Russell2020AI} or Monte Carlo tree search~\cite{Browne2012MCTS} is a natural direction for future work, though simulation cost per node would need to be addressed.
\paragraph{Generalization of the three-level guidance framework.}
Although the system was designed for railway traction unit scheduling, the three-level guidance framework described in Section~\ref{sec:guidance} is likely applicable to similar semi-automatic, human-in-the-loop workflows in which candidate actions can be grouped, ranked by a primary metric, and evaluated using additional contextual information before committing to modify.
Potential application domains include crew scheduling, vehicle routing, and production planning, where similar human-in-the-loop refinement patterns arise from optimization outputs.
From a visualization design perspective, the key differentiator from prior simulation steering work is that discrete combinatorial modification spaces do not admit gradient-based navigation~-- making explicit ranking and progressive disclosure a principled necessity rather than a convenience.
\paragraph{Scalability.}
The seven-day planning horizon is fixed by the planning workflow, so scalability is primarily a question of the number of traction units in the cluster, which typically ranges from a handful to approximately one hundred, with 15 to 20 units representing a medium to large case.
For very large clusters, adaptive aggregation strategies~-- for example, clustering nearby crossing candidates in the overview~-- could help manage the increasing visual density and number of candidates.
\section{Summary and Conclusion}
\label{sec:conclusion}
We present a visual analytics system for the interactive exploration and refinement of railway traction unit circulation plans. 
The system integrates a domain-familiar circulation plan view with simulation-based evaluation, allowing experts to assess the operational robustness of a planned schedule and iteratively improve it through crossings.

A central contribution of the system is the three-level guidance mechanism, which supports the identification and evaluation of crossing candidates at three progressively detailed levels: a spatial overview of aggregated candidates ranked by a primary KPI, a local comparison of candidates within a selected location, and a full per-candidate
breakdown including secondary attributes. 
Although instantiated here for railway scheduling, the mechanism is designed to generalize to other iterative modification workflows in which candidate actions can be grouped, ranked by a primary metric, and evaluated using additional contextual information.

The provenance tree supports non-linear exploration by recording every schedule state produced during a session, allowing experts to branch, backtrack, and compare alternative refinement paths. 
The reference overlay in the KPI charts enables direct comparison between any two states, supporting informed decisions about which exploration path to pursue.

The system was developed in close collaboration of visualization and domain experts, and its utility is demonstrated through three use cases covering delay minimization, tandem continuity optimization, and maintenance window balancing, showing that the system supports expert reasoning from KPI-guided modifications to operationally motivated decisions that trade off one metric against another.
%
%\newpage
\section{Supplemental Material Instructions}
\label{sec:supplement_inst}
%
%The full paper, including all supplemental materials, is available on arXiv at \url{XXX}.
The appendix (Sect.~\ref{suppl:prevworkflow}) provides more information on \ref{sec:background} Background~/ Railway
Scheduling and includes an additional Figure (Fig. \ref{fig:legacy_circulationplan}). %All supplemental figures are referred to in the main text.
%
% %
% If we decide to add appendix or supplemental material in form of additional images we should mention it here!

% In support of transparent research practices and long-term open science goals, you are encouraged to make your supplemental materials available on a publicly-accessible repository.
% Please describe the available supplemental materials in the \hyperref[sec:supplemental_materials]{Supplemental Materials} section.
% These details could include (1) what materials are available, (2) where they are hosted, and (3) any necessary omissions.
%
%% if specified like this the section will be omitted in review mode
%

\acknowledgments{%
The VRVis GmbH is funded by BMIMI, BMWET, Tyrol, Vorarlberg and Vienna Business Agency in the scope of COMET - Competence Centers for Excellent Technologies (911654) which is managed by FFG. Parts of this work have been done at CEDAS, the Center for Data Science at the Univ.\ of Bergen, Norway.  
}
\bibliographystyle{abbrv-doi-hyperref}
\bibliography{LEO}
\newpage
\appendix
\onecolumn\large 
\begin{center}\Large Guided Exploration of Iterative Schedule Modifications:\\A Design Study on Railway Traction Unit Scheduling\end{center}
\begin{center}\huge{}Supplementary Material\end{center}
\vspace{2ex}\section{Supplementary Material: Background/Railway Scheduling}\label{suppl:prevworkflow}

\par\bigskip\noindent 
This supplementary section provides additional information to support the content presented in Section \ref{sec:background} Background/Railway Scheduling.
\setlength{\footnotesep}{4mm}
\subsection{Current Workflow of Traction Unit Scheduling}
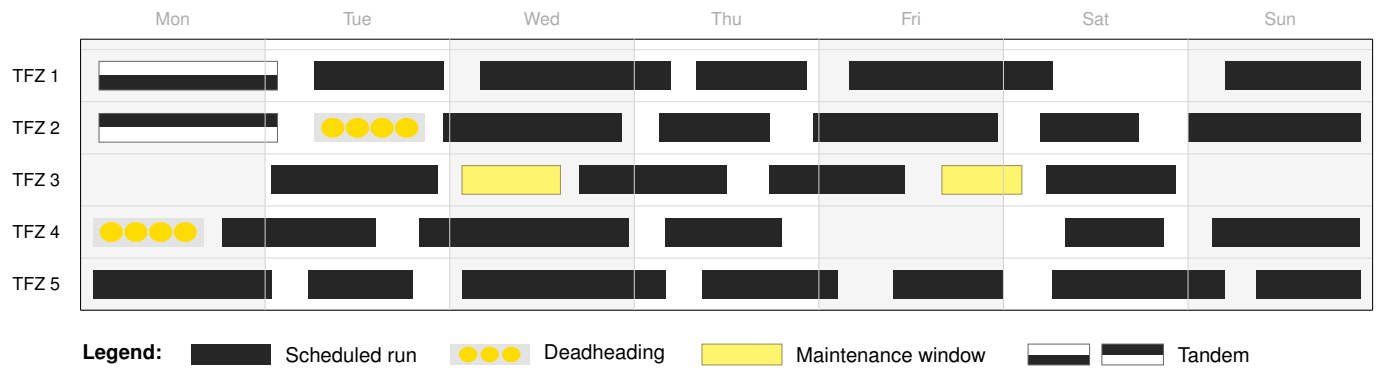
\begin{figure}[t!]
\centering
\resizebox{\textwidth}{!}{%
\begin{tikzpicture}[x=0.85cm, y=0.72cm]

% Parameters
\def\nrows{5}
\def\daywidth{3.0}
\def\ndays{7}
\def\totalwidth{21}
\def\rh{0.55}% bar half-height

% Alternating day shading -- clipped to plan height
\foreach \d in {0,2,4,6} {
    \fill[gray!8] (\d*\daywidth,-\nrows)
        rectangle (\d*\daywidth+\daywidth,0.2);
}

% Day labels
\foreach \d/\lbl in {0/Mon,1/Tue,2/Wed,3/Thu,4/Fri,5/Sat,6/Sun}{
    \node[font=\small\sffamily,text=gray!65]
        at (\d*\daywidth+\daywidth/2, 0.6) {\lbl};
}

% Row labels and separators
\foreach \r/\lbl in {1/TFZ 1,2/TFZ 2,3/TFZ 3,4/TFZ 4,5/TFZ 5}{
    \node[anchor=east,font=\small\sffamily]
        at (-0.2,-\r+0.5) {\lbl};
    \draw[gray!30,thin](0,-\r+1)--(\totalwidth,-\r+1);
}
\draw[gray!30,thin](0,-\nrows)--(\totalwidth,-\nrows);

% -------------------------------------------------------
% ROW 1: TFZ 1 -- remotely controlled tandem partner
% -------------------------------------------------------
\fill[white]    (0.3,-1+\rh/2+0.5) rectangle (3.2,-1+0.5);
\fill[black!85] (0.3,-1+0.5)       rectangle (3.2,-1-\rh/2+0.5);
\draw[black!60,thin](0.3,-1+\rh/2+0.5) rectangle (3.2,-1-\rh/2+0.5);
% Normal runs -- varied lengths, varied gaps
\fill[black!85](3.8,-1+\rh/2+0.5) rectangle (5.9,-1-\rh/2+0.5);
\fill[black!85](6.5,-1+\rh/2+0.5) rectangle (9.6,-1-\rh/2+0.5);
\fill[black!85](10.0,-1+\rh/2+0.5) rectangle (11.8,-1-\rh/2+0.5);
\fill[black!85](12.5,-1+\rh/2+0.5) rectangle (15.8,-1-\rh/2+0.5);
%\fill[black!85](16.3,-1+\rh/2+0.5) rectangle (17.9,-1-\rh/2+0.5);
\fill[black!85](18.6,-1+\rh/2+0.5) rectangle (20.8,-1-\rh/2+0.5);

% -------------------------------------------------------
% ROW 2: TFZ 2 -- master tandem + deadheading + normal runs
% -------------------------------------------------------
\fill[black!85](0.3,-2+\rh/2+0.5) rectangle (3.2,-2+0.5);
\fill[white]   (0.3,-2+0.5)       rectangle (3.2,-2-\rh/2+0.5);
\draw[black!60,thin](0.3,-2+\rh/2+0.5) rectangle (3.2,-2-\rh/2+0.5);
% Deadheading
\fill[gray!22](3.8,-2+\rh/2+0.5) rectangle (5.6,-2-\rh/2+0.5);
\foreach \x in {4.1,4.5,4.9,5.3}{
    \fill[yellow!80!orange](\x,-2+0.5) circle(0.19);
}
% Normal runs -- varied
\fill[black!85](5.9,-2+\rh/2+0.5) rectangle (8.8,-2-\rh/2+0.5);
\fill[black!85](9.4,-2+\rh/2+0.5) rectangle (11.2,-2-\rh/2+0.5);
\fill[black!85](11.9,-2+\rh/2+0.5) rectangle (14.9,-2-\rh/2+0.5);
\fill[black!85](15.6,-2+\rh/2+0.5) rectangle (17.2,-2-\rh/2+0.5);
\fill[black!85](18.0,-2+\rh/2+0.5) rectangle (20.8,-2-\rh/2+0.5);

% -------------------------------------------------------
% ROW 3: TFZ 3 -- two maintenance windows + normal runs
% -------------------------------------------------------
%\fill[black!85](0.2,-3+\rh/2+0.5) rectangle (2.4,-3-\rh/2+0.5);
\fill[black!85](3.1,-3+\rh/2+0.5) rectangle (5.8,-3-\rh/2+0.5);
% First maintenance window (Wed)
\fill[yellow!70](6.2,-3+\rh/2+0.5) rectangle (7.8,-3-\rh/2+0.5);
\draw[yellow!50!black,thin](6.2,-3+\rh/2+0.5) rectangle (7.8,-3-\rh/2+0.5);
\fill[black!85](8.1,-3+\rh/2+0.5) rectangle (10.5,-3-\rh/2+0.5);
\fill[black!85](11.2,-3+\rh/2+0.5) rectangle (13.4,-3-\rh/2+0.5);
% Second maintenance window (Sat)
\fill[yellow!70](14.0,-3+\rh/2+0.5) rectangle (15.3,-3-\rh/2+0.5);
\draw[yellow!50!black,thin](14.0,-3+\rh/2+0.5) rectangle (15.3,-3-\rh/2+0.5);
\fill[black!85](15.7,-3+\rh/2+0.5) rectangle (17.8,-3-\rh/2+0.5);
%\fill[black!85](18.3,-3+\rh/2+0.5) rectangle (20.7,-3-\rh/2+0.5);

% -------------------------------------------------------
% ROW 4: TFZ 4 -- deadheading start + normal runs
% -------------------------------------------------------
\fill[gray!22](0.2,-4+\rh/2+0.5) rectangle (2.0,-4-\rh/2+0.5);
\foreach \x in {0.5,0.9,1.3,1.7}{
    \fill[yellow!80!orange](\x,-4+0.5) circle(0.19);
}
\fill[black!85](2.3,-4+\rh/2+0.5) rectangle (4.8,-4-\rh/2+0.5);
\fill[black!85](5.5,-4+\rh/2+0.5) rectangle (8.9,-4-\rh/2+0.5);
\fill[black!85](9.5,-4+\rh/2+0.5) rectangle (11.4,-4-\rh/2+0.5);
%\fill[black!85](12.2,-4+\rh/2+0.5) rectangle (15.4,-4-\rh/2+0.5);
\fill[black!85](16.0,-4+\rh/2+0.5) rectangle (17.6,-4-\rh/2+0.5);
\fill[black!85](18.4,-4+\rh/2+0.5) rectangle (20.8,-4-\rh/2+0.5);

% -------------------------------------------------------
% ROW 5: TFZ 5 -- normal runs, varied lengths
% -------------------------------------------------------
\fill[black!85](0.2,-5+\rh/2+0.5) rectangle (3.1,-5-\rh/2+0.5);
\fill[black!85](3.7,-5+\rh/2+0.5) rectangle (5.4,-5-\rh/2+0.5);
\fill[black!85](6.2,-5+\rh/2+0.5) rectangle (9.5,-5-\rh/2+0.5);
\fill[black!85](10.1,-5+\rh/2+0.5) rectangle (12.3,-5-\rh/2+0.5);
\fill[black!85](13.2,-5+\rh/2+0.5) rectangle (15.0,-5-\rh/2+0.5);
\fill[black!85](15.8,-5+\rh/2+0.5) rectangle (18.6,-5-\rh/2+0.5);
\fill[black!85](19.1,-5+\rh/2+0.5) rectangle (20.8,-5-\rh/2+0.5);

% -------------------------------------------------------
% Borders and day dividers (drawn on top of shading)
% -------------------------------------------------------
\draw[black,thin](0,0.2)--(\totalwidth,0.2);
\draw[black,thin](0,-\nrows)--(\totalwidth,-\nrows);
\draw[black,thin](0,0.2)--(0,-\nrows);
\draw[black,thin](\totalwidth,0.2)--(\totalwidth,-\nrows);
\foreach \d in {1,...,6}{
    \draw[gray!35,thin](\d*\daywidth,0.2)--(\d*\daywidth,-\nrows);
}

% -------------------------------------------------------
% Legend -- single row
% -------------------------------------------------------

\node[anchor=west,font=\footnotesize\sffamily\bfseries]
    at(-0.1,-\nrows-0.85){Legend:};
% Scheduled run  (starts at 1.8)
\fill[black!85](1.8,-\nrows-1.05) rectangle (3.1,-\nrows-0.65);
\node[anchor=west,font=\footnotesize\sffamily]
    at(3.2,-\nrows-0.85){Scheduled run};
% Deadheading  (3.2 + 2.3 + 0.5 = 6.0)
\fill[gray!22](6.0,-\nrows-1.05) rectangle (7.3,-\nrows-0.65);
\foreach \x in {6.3,6.65,7.0}{
    \fill[yellow!80!orange](\x,-\nrows-0.85) circle(0.15);
}
\node[anchor=west,font=\footnotesize\sffamily]
    at(7.4,-\nrows-0.85){Deadheading};
% Maintenance window  (7.4 + 1.9 + 0.5 = 9.8)
\fill[yellow!70](10.1,-\nrows-1.05) rectangle (11.4,-\nrows-0.65);
\draw[yellow!50!black,thin](10.1,-\nrows-1.05) rectangle (11.4,-\nrows-0.65);
\node[anchor=west,font=\footnotesize\sffamily]
    at(11.5,-\nrows-0.85){Maintenance window};
% Tandem  (11.2 + 3.2 + 0.5 = 14.9)
\fill[black!85](15.4,-\nrows-1.05) rectangle (16.4,-\nrows-0.65);
\fill[white]   (15.4,-\nrows-0.85) rectangle (16.4,-\nrows-0.65);
\draw[black!60,thin](15.4,-\nrows-1.05) rectangle (16.4,-\nrows-0.65);
\fill[white]   (16.6,-\nrows-1.05) rectangle (17.6,-\nrows-0.65);
\fill[black!85](16.6,-\nrows-0.85) rectangle (17.6,-\nrows-0.65);
\draw[black!60,thin](16.6,-\nrows-1.05) rectangle (17.6,-\nrows-0.65);
\node[anchor=west,font=\footnotesize\sffamily]
    at(17.7,-\nrows-0.85){Tandem};
    
\end{tikzpicture}%
}% end resizebox
\caption{Schematic illustration of the domain-standard traction unit
circulation plan format~\cite{paprer2025rolling} as used in planning
practice and adopted as the basis for the Circulation Plan View in our
system. Each row represents one traction unit; horizontal bars span
scheduled train runs over the reference week. TFZ~1 and TFZ~2 form
a tandem pair: TFZ~2 is the master unit (black bar on top, white on
bottom) and TFZ~1 is the remotely controlled unit (white on top, black
on bottom). Deadheading runs are shown as bars with yellow circles,
and maintenance windows as yellow bars. This schematic was recreated
from the domain-standard representation; screenshots of the proprietary
planning software used by our domain expert co-authors cannot be
shared due to confidentiality constraints on operational data.}
\label{fig:legacy_circulationplan}
\end{figure}

The current workflow of creating traction unit schedules is heavily contextual. Technical capabilities, customer requirements, predictability of the traffic and so forth, all play a significant role during the process design; an abstract depiction could however follow these steps:
\begin{enumerate}
\item Tendering/Transport Planning/ Timetabling:
Possible traction unit circulations are often already taken into account during the early planning stages. Efficient circulations are being considered when planning the virtual trains themselves to ensure overall efficiency. 

\item Traction Unit Assignment:
In this step, the decision is made which trains are served by which type of traction unit. In academia as well as in real life the distinction to step 1 and 3 can be both vague and fluent. Methodologically this is a complex task that mostly revolves around meeting requirements such as electrification, train security systems, local certifications, tractional force in combination with track elevation, passenger or freight requirements etc.

\item Traction Unit Scheduling:
In this step the actual circulation plans (as discussed in this paper) are created. A schematic illustration of the domain-standard circulation plan is shown in Figure \ref{fig:legacy_circulationplan}. A variation of such a plan is used by planners and researchers in railway scheduling \cite{paprer2025rolling}. If a schedule is not circular, deadheading trips are added. During the manual creation or manipulation, planners rely on simple operations such as swapping, shifting, crossing.
\begin{enumerate}
\item[\textbullet] Swapping: Trains or parts of trains are swapped between two traction units.
\item[\textbullet] Shifting: Trains or parts of trains are shifted from one traction unit to another.
\item[\textbullet] Crossing: The entire subsequent schedules of two traction units are exchanged at a common point in time and space.
\end{enumerate}
\end{enumerate}

\noindent These operations are digitally supported through comparative tables and intuitive icons, yet the combinatorial complexity can be overwhelming. Hence, the guidance system as proposed in this paper has the potential to support planners during this step.    

\subsection{Legacy Design Details}

In this paper several interface design details are based on designs currently used by planners and planning tools of the project partner. They can hence be regarded as legacy artifacts; which includes but is not limited to the following:
\begin{itemize}
    \item Deadheading: A horizontal bar with yellow circles directly corresponds to the current form of visualization of this task.
    \item Crossing: Crossing indicators appear as gray crosses in the Circulation Plan View, corresponding to the symbol on the icon that planners use to open the crossing-interface. Directly indicating crossing opportunities within the plan is novel.
    \item  Multiple simultaneous crossings: Simultaneous crossings between multiple traction units (e.g., at major hubs) are shown as individual crosses, each with a subscript denoting the identifier of the respective traction units available for crossing.
    \item Horizontally split bars: The horizontal split in certain bars indicates that multiple traction units are assigned to a train. If the top part is solid, then this unit is in the lead. 
\end{itemize}

\noindent The spatial arrangement of the four main panes reflects deliberate design decisions made in consultation with the domain expert co-authors. 
The Circulation Plan View occupies the central and largest area on the left, consistent with its role as the primary analysis surface and the representation most familiar to planning practitioners~-- who naturally expected it to be the dominant element. 
The KPI charts are placed to the upper right, immediately adjacent to the plan rows they correspond to. 
The Circulation Plan Overview occupies the lower right, a position consistent with the conventional role of overview maps in other tools (such as minimaps in games and geographic applications), where they serve as secondary navigation aids rather than primary analysis surfaces. 
The Provenance and Favorites panel is placed between the KPI charts and the overview, reflecting the workflow sequence: experts first analyze the plan and KPI charts, then use the provenance tree to manage explored states, and finally consult the overview to identify new crossing candidates. 
We note that different arrangements are certainly possible in other application domains; the current layout was chosen by and validated with the domain experts for this specific planning context.

% ============================================================
% TikZ schematic v2: standard traction unit circulation plan
% For Appendix A.2 (Legacy Design Details)
% Place at top of page 2 of appendix, after bullet list, before A.3
% Requires: \usepackage{tikz} in preamble
% ============================================================

\subsection{Expert Description} \label{suppl:Experts}
This paper, as well as the proposed visual interface, was developed in close cooperation with two domain experts, who co-authored the paper:
\begin{itemize}
    \item Expert A: 5+ years of domain specific work experience (locomotive scheduling). Focus on long-term planning. Product owner of a digital planning support tool.
    \item Expert B: 3+ years of domain specific work experience (locomotive re-scheduling). Focus on short-term planning and stochasticity. Product owner of a digital planning support tool.
\end{itemize}

\end{document}